\documentclass[sigconf, screen, nonacm]{acmart}

\usepackage{booktabs}
\usepackage{longtable}

\usepackage[colorinlistoftodos,disable]{todonotes}
\usepackage{soul}

\AtBeginDocument{%
  }

\begin{document}

\title[Distributed Seed Storage]{Share a Tiny Space of Your Freezer to Increase Resilience of Ex-situ Seed Conservation}


\author{Andrea Vitaletti}
\email{vitaletti@diag.uniroma1.it}
\orcid{1234-5678-9012}
\affiliation{%
  \institution{Sapienza University of Rome}
  \country{Italy}
}


\begin{abstract}
More than 95\% of the crop genetic erosion articles analyzed in \cite{https://doi.org/10.1111/nph.17733} reported changes in diversity, with nearly 80\% providing evidence of loss. 
The lack of diversity presents a severe risk to the security of global food systems. Without seed diversity, it is difficult for plants to adapt to pests, diseases, and changing climate conditions. Genebanks, such as the Svalbard Global Seed Vault, are valuable initiatives to preserve seed diversity in a single secure and safe place. However, according to our analysis of the data available in the Seed Portal, the redundancy for some species might be limited, posing a potential threat to their future availability. Interestingly, the conditions to properly store seeds in genebanks, are the ones available in the freezers of our homes. This paper lays out a vision for Distributed Seed Storage relying on a peer-to-peer infrastructure of domestic freezers to increase the overall availability of seeds.  We present a Proof-of-Concept focused on monitoring the proper seed storing conditions and incentive user participation through a Blockchain lottery. The PoC proves the feasibility of the proposed approach and outlines the main technical issues that still need to be efficiently solved to realize a fully-fledged solution.
\end{abstract}

\maketitle


\section{Introduction}

\todo[inline]{\st{The 75\% loss of agricultural biodiversity is an old, unconfirmed number}}
\todo[inline]{\st{Svalbard is not an active genebank to be compared with MSB. It serves a different purpose}}
\todo[inline]{\st{There is no mention of the Multilateral System (MLS)} and no discussion on the policy implications of this concept}
\todo[inline]{Seed conservation is not just about storing seeds in the freezer - like a museum. This is again an old concept. There is a workflow to ensure genetic integrity and the long-term availability and use of germplasm, including documentation, characterization, evaluation, viability testing, health testing, regeneration, multiplication, distribution, etc.}

There are over 50,000 edible plants worldwide, but just 15 provide 90\% of the world’s food energy intake and among these, rice,  corn (maize), and wheat contribute up to two-thirds \footnote{\url{https://education.nationalgeographic.org/resource/food-staple/}}. We have to put  all our efforts to preserve those fundamental energy assets. 

\textbf{The need for seed diversity} A key objective in preserving these plants is to protect seed diversity, that is, the wide range of plant species, cultivars and genetic materials found within a given crop. This genetic diversity is a vital resource on which the future of our food systems depends.

As climate change intensifies, bringing extreme weather events along with emerging pests and diseases, the presence of specific genes becomes critical. These genes enable certain crop varieties to withstand heat, drought, pests, and pathogens, forming the foundation of agricultural resilience and adaptation.

A limited seed pool not only threatens our ability to cope with environmental stresses but also risks the loss of flavor, nutritional value, and the cultural heritage embedded in traditional cuisines. Genetic diversity improves agriculture's ability to adapt to changing climates, regional growing conditions, and evolving consumer preferences.

Without broad access to diverse crop varieties, future generations will face serious constraints in shaping agriculture to meet their needs and respond to environmental challenges.

We do not simply need to preserve cultivated crops, but also their Crop Wild Relatives (CWR). They are wild plant species that are closely related to cultivated crops and can provide genetic diversity that may not be available in current cultivated varieties. This diversity may be used to improve the productivity, resileance, and quality of agricultural products. The Earth is losing plant genetic diversity at an unprecedented rate. More than 95\% of the crop genetic erosion articles analyzed in \cite{https://doi.org/10.1111/nph.17733} reported changes in diversity, with nearly 80\% providing evidence of loss. The extent of this loss varied depending on species, taxonomic and geographic scale, region, and analytical approach. In many countries, seed diversity has been irreversibly diminished, limiting farmers' crop options.

\textbf{Efforts to preserve seed diversity.} Preservation of seed diversity is considered a common fundamental goal among the countries organized in the Multilateral System of Access and Benefit Sharing (MLS \cite{MLS}), part of the International Treaty on Plant Genetic Resources for Food and Agriculture (PGRFA).  MLS allows member states to share and exchange plant genetic resources for research, breeding, conservation, and training. Recipients sign a Standard Material Transfer Agreement (SMTA)~\cite{SMTA} that specifies conditions for use and benefit sharing. MLS includes 64 of the world’s most important crops that together represent 80\% of all human consumption derived from plants~\cite{MLSannex1}.

The main strategies to preserve seed diversity are conservation \emph{in situ} and \emph{ex situ} .

\begin{table*}[ht]
\centering
\begin{tabular}{l|l|l}
\toprule
\textbf{Criteria} & \textbf{In Situ Conservation} & \textbf{Ex Situ Conservation} \\
\midrule
\textbf{Location} & Natural environment / farms & Seed banks, seed vaults, labs \\
\textbf{Evolution} & Plants continue to adapt naturally. & Static; no natural evolution \\
\textbf{Risk of Loss} & Higher (e.g., land-use change, climate) & Lower (secure and controlled conditions) \\
\textbf{Cultural Link} & Strong; maintains traditional knowledge & Weak or absent \\
\textbf{Access \& Sharing} & Local and informal & Global access via formal mechanisms (e.g., SMTA) \\
\textbf{Documentation} & Often limited or informal & Detailed and standardized \\
\textbf{Cost} & Lower short-term cost & Higher cost (infrastructure and maintenance) \\
\textbf{Best Use} & Wild relatives, landraces & Major crops, backup, and global breeding \\
\bottomrule
\end{tabular}
\caption{Comparison between in situ and ex situ seed conservation methods.}
\end{table*}

In situ conservation relies on the conservation of seeds and plants in their natural or traditional environment, such as farms where traditional varieties (landraces) are maintained or in the wild where wild relatives of crops grow. Participatory seed breeding and community seed systems involve local farmers in storing, conserving, sharing and improving diverse varieties adapted to local needs. Seed guardians (or seed custodians) are individuals or communities (e.g. \url{https://www.navdanya.org/}, \url{https://www.connectedseeds.org/}) often small farmers, indigenous people \url{https://www.wfp.org/stories/supporting-latin-americas-custodians-seeds}, or members of community seed bank (e.g. \url{https://seedsavers.org/}), who actively preserve, cultivate, and exchange traditional or locally adapted crop varieties, often outside of formal, industrial seed systems .

Although in situ conservation has the merit of actively participating local communities and farmers, thus contributing to the conservation of not only the seeds, but also the traditions and skills to take care of them, the access to the material and the documentation is fragmented and difficult to handle. These problems have led to the development of several more structured ex situ conservation initiatives, where seeds are preserved outside their natural habitat in the so-called \emph{seed banks}. 

\textbf{Seed banks.} A seed bank is a gene bank, that stores seeds to preserve genetic diversity, namely the genes that make each plant variety unique. It ensures that this genetic heritage is safely conserved and available for people to use.  

The Millennium Seed Bank Partnership (MSB) \cite{MSB}
is the largest ex situ plant conservation program in the world, mainly focused on wild species. MSB focuses on the conservation of as many species as possible (interspecific diversity) prioritizing those that are endangered, endemic, or economically important.
It  provides seeds to registered organizations for research, restoration, and reintroduction purposes.

Crop Trust \cite{CroptTrust} is an international organization focused on building and supporting a global system of seed banks to preserve the genetic diversity of food crops, especially cultivated varieties and landraces. It provides long-term grants and partnership agreements to several international seed banks, particularly those within the Consultative Group on International Agricultural Research (CGIAR) system. CGIAR genebanks~\cite{CGIAR} manage collections of more than 20 staple crops in 10 locations across five continents to distribute seeds for research, training, and breeding under the MLS framework. 

Genesys is an online platform\footnote{\url{https://www.genesys-pgr.org/}} managed bt Crop Trust that provides access to information about plant genetic resources conserved in genebanks around the world. It serves as a global catalogue, aggregating data on millions of accessions from over 450 genebanks, included CGIAR ones.

The Crop Trust also co-manages the Svalbard Global Seed Vault in Norway (SGSV)\footnote{\url{https://www.seedvault.no/}}, which serves as a global backup facility for seed banks worldwide~\cite{Hopkin2008}. 
Its purpose is to provide long-term backing for genebank collections to secure the foundation of our future food supply against threats such as mismanagement, accidents, equipment failures, funding cuts, war, sabotage, disease, and natural disasters. In 2019, seeds were returned from Svalbard to restock an international genebank destroyed by the war in Syria. 

Contrary to MSB it is not open to researchers and the depositors are the only ones who can withdraw their own seeds. The focus of SGSV is to preserve as much diversity as possible within crop species and their wild relatives (intraspecific diversity). 

Seed banks must ensure that their holdings remain alive. They periodically check the viability of the samples, namely their ability to grow into productive plants and produce fresh material when needed. However, the fundamental strategy to preserve such diversity is the availability of good backup systems, namely other seed banks where duplicate samples are delivered for safe keeping. 

\textbf{Open problem.} The SGSV is unanimously considered the backup seed facility that provides the highest standards of safety and security. The location where SGSV is built lacks tectonic activity and has permafrost, which aids in preservation; furthermore, it is 130 m above sea level, keeping the site dry even if the ice caps melt.  The refrigeration units also cool the seeds to -18,-20 $^{\circ}$C, the internationally recommended standard temperature~\cite{faostand}. If the equipment fails, it is estimated that it will take two centuries for it to warm to 0 $^{\circ}$C. 

The 2023 SGSV Annual Progress Report~\cite{repSGSV2023} reveals that, 1267127 samples deposited by more than 100 genebanks/institutes and representing more than 6,000 plant species are stored in the facility. The SGSV now safeguards seeds of more than 250,000 types of wheat, 160,000 types of rice and 46,000 types of maize, the fundamental sources of food energy intake. 

\begin{figure}[h]
    \centering
    \includegraphics[width=0.7\linewidth]{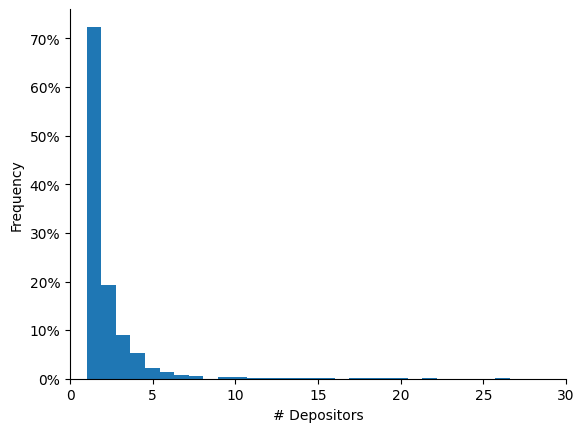}
    \caption{The distribution of the depositors for the species in the SGSV. The majority have a single depositor posing a potential threat to seed availability. }
    \label{fig:dist}
\end{figure}
\smallskip
Figure \ref{fig:dist} shows the distribution of the depositors (i.e. genebanks and institutes) for the species obtained by analyzing the data from the SGSV Seed Portal~\footnote{\url{https://seedvault.nordgen.org/Search}}.

The vast majority of species have a single depositor. In other words, the data show that the samples are available at SGCV and at the depositor only. As an example, Oryza sativa (i.e. the Asian cultivated rice) is the most common rice cultivated as a cereal. The result of the query to the Seed Portal with the keyword \emph{Species: Oryza sativa} is 29 depositors, 171193 accessions and 136 countries of collection. The same query with the keyword \emph{Species: Hygroryza aristata}, gives 1 depositor (i.e. the Rice Research Institute), 4 accessions and 1 country of collection as also confirmed by the Genesys online platform.
Rice is currently by far more ``important'' than Hygroryza for food supply, however, the Hygroryza is a crop wild relative (CWR) of rice~\cite{JACQUEMIN2013147} and, as already observed, it might contribute to the development of more resilient new rice varieties.


The low level of redundancy in the availability of seeds samples backups for the majority of the species, might have serious consequences; even in the best genebanks, things can occasionally go disastrously wrong. The question we aim at investigating in this paper is whether it is possible to further increase the overall availability of seeds by a robust and resilient backup infrastructure capable of preserving seed diversity with limited investments. 


\textbf{Contributions of the paper.} Interestingly, the temperature conditions in domestic freezers can meet the international standard for long-term seed conservation \cite{faostand,faostandpract}. The main idea of this paper is to develop a system inspired by the principles of the shared economy, capable of safely storing the seeds in our home freezers, thus contributing to the development of a Distributed Seed Storage (DSS) that can increase the availability of seeds and the overall resilience of the world’s crop diversity. To our knowledge, this paper is the first attempt to build a peer-to-peer infrastructure to preserve seed diversity. We first identified a set of requirements, and then we built a proof-of-concept of the DSS that proves the technical feasibility of two fundamental tasks, namely the IoT monitoring activity of seed conservation conditions in domestic freezers and the interaction of the IoT node with a Blockchain smart contract to autonomously manage a lottery to incentive users participation. The development of this PoC, allows us to identify the main issues that need to be investigated in future work.

\section{Background}

Conservation of orthodox seeds in seed genebanks involves several operations~\cite{faostand,faostandpract}: acquisition, drying
and storage, seed viability monitoring and regeneration, characterization
and evaluation, documentation and distribution and safety duplication. In this paper we limit ourselves to considering the storage. Although this activity is only one of the necessary operations, the existence of genebanks, and in particular the ones specifically designed as backup facilities such as SGSV, only makes sense if seed are store at proper conditions in safe redundant locations.  

\textbf{Conditions to preserve seeds.} Moisture, temperature, and the proportion of oxygen are key environmental factors that affect seed deterioration and loss of viability~\cite{https://doi.org/10.1111/rec.13174}. Reducing seed moisture content (MC) to certain thresholds increases longevity predictably for approximately 90\% of species. These species are classified as being \emph{orthodox} in their seed storage requirements, and generally retain viability and germinability even after storage for long periods under suitably dry, cool conditions. In general, most of our food energy intake from plants is orthodox.
The Food and Agriculture Organization (FAO) standardized the process for long-term storing of orthodox seeds~\cite{faostand,faostandpract}. We require our system to comply with these standards, as specified in \textbf{R1}, section \ref{sec:req}.


\textbf{IoT and Shared Economy.} We share the definition of shared economy by Frenken et al. \emph{``consumers granting each other temporary access to under-utilized physical assets (\emph{idle capacity}), possibly for money.''}~\cite{FRENKEN20173}. 

Example of shared economy ranges from carpooling (e.g. BlaBlaCar), to share rooms in an apartment (e.g. Airbnb) or to  utilize transport capacity that may be wasted in inefficient transport operations (e.g. Shiply), just to mention few examples. In our case, the idle capacity is a tiny portion of the freezer sufficient to store the seeds. 

The shared economy can benefit from the integration with the Internet of Thing (IoT)\cite{LANGLEY2021853}. IoT can be used to monitor the availability of the idle capacity and to verify that some requirements are meet (see for example \textbf{R1} section \ref{sec:req}). 






\textbf{Encourage Users Participation by Incentives.} The success of this project depends on the active participation of a sufficiently high number of users. The nature of the project, and the limited freezer space required to store the seeds, give us hope for a proactive and numerous participation. However, suitable incentives might clearly help. Indeed, as discussed in  
\cite{https://doi.org/10.1002/asi.23552}, enjoyment, economic incentive, reputation, and self-fulfillment are the main motivations behind the participation to the shared economy. In particular, we are interested in evaluating how the Blockchain technology can distribute such incentives.  In~\cite{10.1145/3539604}, the authors discuss Blockchain-based incentive mechanisms. In this paper the participation to a lottery is the incentive to participate: only peers actively contributing  to the DSS can buy tickets and hope to win.  



\section{Requirements}\label{sec:req}
In this section we describe the main requirements that drive the design of the DSS proof-of-concept. 
\begin{enumerate}
    \item[\textbf{R1}] Storing conditions of seeds in domestic freezers must comply with \cite{faostand,faostandpract}. In this paper, we focus on long-term storage of orthodox seeds. Specifically we assume that after drying, samples meant for long-term storage are packaged under controlled conditions, in clearly labelled airtight containers (Standard 4.2.2 \cite{faostand}) and then samples  are ideally stored at -18$^{\circ}$C ± 3 percent  and relative humidity of 15 ± 3 percent (Standard 4.2.3 \cite{faostand}). Figure 4 of \cite{faostandpract} is a summary diagram of the workflow and activities for drying and storage and clarifies that the use of subzero freezers is acceptable if ideal conditions are not available.
\end{enumerate}
The success of this project relies on the active and numerous participation of users. To encourage the participation, the system should:
\begin{enumerate}
    \item[\textbf{R2}] Simplify the installation without requiring any specific technical skill or infrastructure. The installation of the system should be user friendly and based on off-the-shelves equipment. We only assume users have a WiFi at home. 
    \item[\textbf{R3}] Not require unnecessary cables which complicates and limit the installation options. 
    \item[\textbf{R4}] Run for at least 1 year without the need of intervention by the user.
    \item[\textbf{R5}] Support the implementation of suitable incentive mechanisms.
\end{enumerate}


\section{Proof-of-concept}

The main purpose of this section is to demonstrate the feasibility of the proposed approach by the implementation of a Proof-of-Concept (PoC) of the Distributed Seed Storage driven by the above requirements. The PoC will also  outline the main technical problems that still need to be efficiently solved to realize a fully fledged solution. 

The airtight container storing the seeds is equipped with a humidity and temperature sensor (the DHT22) connected to an IoT sensor node, in our case the Esp32 Devkit V1 (Esp32 in the following), which implements the logic to acquire the sensor readings and to deliver them to the smart contract in charge of dispensing the incentives to the participants. 

The Esp32 delivers the data to the smart contract by the on-board WiFi. WiFi access points are nowadays largely available in homes (see \textbf{R2}) making this technology appropriate for the project, 
furthermore, the nature of the phenomenon to be monitored does not require frequent data acquisition; temperature and humidity in the freezer do not suddenly change.  For this reason, the node alternates long period of \emph{deep sleep} with a short active period where two main tasks are performed, namely data acquisition and communication. In \emph{deep sleep} mode, only the RTC Timer and RTC Memory are active, reducing the consumption to 10$\mu$A. The resulting low duty cycle prolongs the lifetime of the node, and allows us to meet the requirement \textbf{R4} as quantified in section \ref{sec:energy}.

To satisfy \textbf{R3}, the Esp32 is powered by a 3.6V LiFePo4. This solution does not require  any voltage regulator that would introduce losses. In the future, we plan to design a more efficient custom board without all unnecessary components. 

The ESP32 operating temperature range\footnote{\url{https://www.espressif.com/sites/default/files/documentation/esp32_datasheet_en.pdf}} is from $-40$ to $125$ $^{\circ}$C, so in principle it can also be placed into the freezer together with the DHT22 temperature and humidity sensor, however, the wifi connectivity could be severely hindered. In our experiments, when the ESP32 is placed into the freezer, more then 40\% of the packets do not reach the server. Furthermore, it is well known the the efficiency of batteries  decreases significantly in cold temperatures: at -20°C to -40°C, a LiFePo4 battery may only achieve about 60\% to 40\% of its rated capacity\footnote{\url{https://www.evlithium.com/Blog/lifepo4-battery-temperature-range-capacity-voltage.html}}. Finally, despite being small, the ESP32 steal some further space in the user's freezer. 
For these reasons, the DHT22 is placed into the freezer and it is connected to the ESP32 placed outside, by three tiny wires. This solution is not fully consistent with \textbf{R3}, but it is acceptable and the arguments discussed above togheter with the simplicity of the installation, makes it compliant with \textbf{R2}.

\begin{figure*}[t!]
    \centering
    \includegraphics[width=0.75\linewidth]{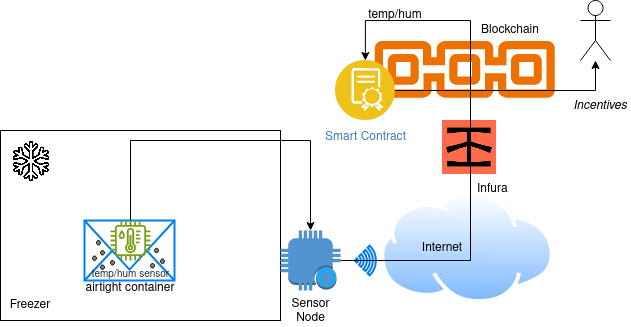}
    \caption{The architecture of the Proof-of-Concept. A temperature and humidity sensor is placed into the freezer and it is connected to an IoT sensor node placed outside. The node communicate the reading to a smart contract in charge of dispensing incentives to encourage users' participation.}
    \label{fig:arch}
\end{figure*}

A sketch of the architecture of the Proof-of-Concept is shown in figure~\ref{fig:arch}. The PoC is build with PlatformIO~\footnote{\url{https://platformio.org/}} and the code is available on \emph{\url{https://github.com/andreavitaletti/PlatformIO/tree/main/Projects/web3E_SC}}.

In the following, we discuss the implementation of the two main tasks performed by the PoC during the active period, namely monitoring the correct storage of seeds and delivering the observed data to the smart contract in charge of incentive users participation.

\subsection{Monitoring the Correct Storage of Seeds}
The temperature and humidity conditions at which seeds are stored in the freezer, are monitored by a DHT22 sensor (see \textbf{R1}). It features excellent technical specifications also in terms of power consumption. The sensor can be queried by a single wire protocol: when the sensor node sends the start signal, the DHT22 change from stand-by-mode (40 $\mu$A)  to measuring-mode (1 mA). The DHT22 will get back to stand-by-mode again upon the completion of the data acquisition.

\begin{figure}[ht!]
    \centering
    \includegraphics[width=1\linewidth]{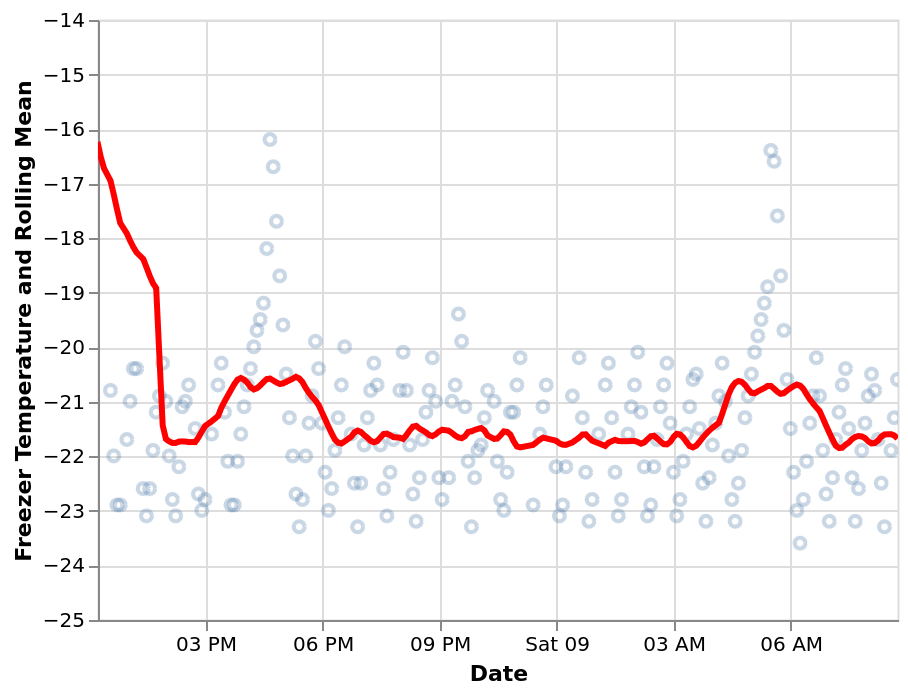}
    \caption{The observed temperature drops immediately after the placement of the sensor into the freezers, then fluctuates noticeably.}
    \label{fig:dht-observation}
\end{figure}

Figure \ref{fig:dht-observation} shows the results of a monitoring activity  on a domestic freezer (BOSCH KGN39X23)  set at -18 $^{\circ}$C of temperature. Samples are taken every 5 minutes. After an initial transient period, the observed temperature oscillates significantly. These conditions do not fulfill the Standard 4.2.3~\cite{faostand} but are consistent with the acceptable subzero conditions indicated in \cite{faostandpract} when the ideal ones are not available. The implications of these conditions in seed conservation and the possibility to reach the ideal ones in domestic freezer are object of future studies. 

Domestic freezers are occasionally opened to access the frozen products. We performed a simple experiment to evaluate the impact of this activity on the temperature. We left the freezer close for 1 hour and then we opened it for 30 secs., a more than sufficient interval of time to access the frozen products; the impact on the temperature is marginal at the first decimal figure of the temperature. 


\subsection{Delivering the observed data to the smart contract}

Once data have been acquired, they are delivered to the blockchain smart contract in charge of distributing the incentives to encourage users' participation (see \textbf{R5}). In our PoC, we envision the development of a blockchain lottery, where users can participate (and possibly have higher chances of winning)  if and only if they are active peers of the DSS. The lottery is implemented with a simple smart contract similar to that available at \url{https://github.com/Scofield-Idehen/Lottery_Contract/blob/main/Lot_Contract.sol}. 
The function \emph{enter(uint temp, uint hum)} allows the participation only of active DSS users, namely users that provide  data on temperature and humidity in the acceptable range. The function \emph{fund}, allows anyone to fund the lottery. The code of the smart contract, together with some sample transactions, can be accessed at \url{https://sepolia.etherscan.io/address/0x19aeacb63eba19e5e159a5870fa6afce5d3b37ec}.

A more complex lottery and a through study on the most suitable incentive mechanisms to be implemented in the smart contract is beyond the scope of this paper and will be investigated in future work. 


Most solutions interfacing  IoT nodes with smart contracts on the Blockchain, relays on gateways. This would greatly simplify the integration, but  we cannot expect that all the houses have the necessary skills and tools to deploy such a relatively complex infrastructure (see \textbf{R2}). For this reason, all the code to interact with the smart contract through the Infura API\footnote{\url{https://www.infura.io/}} is contained in the firmware of the resource constrained ESP32. This might be a pretty challenging solution due to the limited memory available on the ESP32, indeed our code, based on the WEb3E library~\footnote{\url{https://github.com/AlphaWallet/Web3E}}, occupies 93\% of the available flash. This solution foresees only outgoing transmissions (from the home to the Internet), thus increasing the security against possible network threats. 

DHT22 readings are given as floating numbers that are not supported by Solidity. To simplify the management of readings by the smart contract also for negative values, we applied the following simple formula $ 40 + rount(temp) $ with $temp$ ranging from  -40°C to 80°C,  as defined in the technical specification of the DHT22. Consequently, the value transmitted to the smart contract is a single byte representing a value between 0 (-40°C) and 120 (80°C). The transactions can be accessed at the same URL of the smart contract. Floats can be also encoded in 4 bytes using the IEEE 754 standard, but this makes more complex the logic of the smart contract.

\subsection{Energy Consumption}\label{sec:energy}



In this section we report on the experiments  to evaluate the  energy consumption through an INA219\footnote{\url{https://www.ti.com/lit/ds/symlink/ina219.pdf}}. The Esp32 devkit V1, has an always on red led, with a fix consumption of 10mA. We unsoldered the led and the consumption dropped to less than 4mA, however still pretty far from the  nominal $10 \mu A$ in deep sleep. In the following, to evaluate the potentials of our work,  we assume to achieve the nominal current consumption during the sleeping period. 

\begin{figure}
    \centering
    \includegraphics[width=1\linewidth]{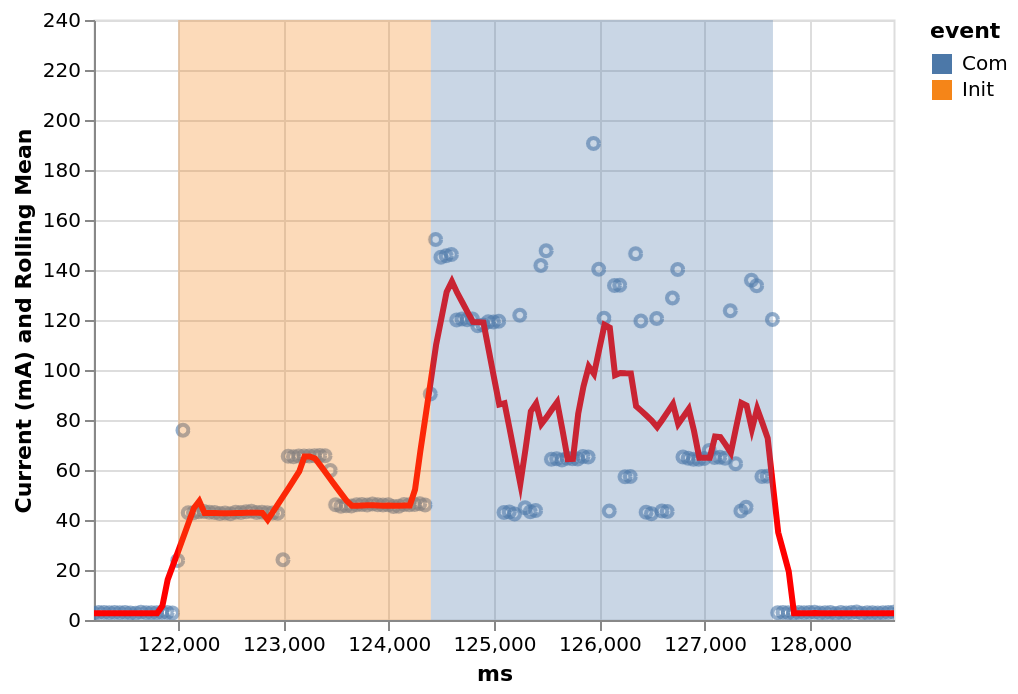}
    \caption{Sampling current consumption with the INA219}
    \label{fig:current}
\end{figure}

We sampled several active periods and figure \ref{fig:current} shows the current consumption in one of those periods;  similar values have been observed for the others. Active periods are made of two tasks, the \emph{init} task where all the necessary components are initialized, and the \emph{com} where the data are read by the DHT and are transmitted to the smart contract.   
The \emph{init} task  lasts about  2400ms ($\approx$0.0007h) 
at about 50mA, namely $\approx$ 0.03mAh, while the \emph{com} task lasts about 3160ms ($\approx$0,0009h) 
at about about 100mA, namely $\approx$ 0.09mAh. 
We assume the duration of the sleep period is $X$ hours at $10 \mu A$, namely $0.01X$mAh. The whole consumption for a cycle is $0.03mAh+0.09mAh+0.01XmAh$. The ESP32 is powered by a common 3.6v, 1.5Ah LiFePo4 battery. This guarantees a number of cycles $1500mAh/(0.12+0.01X)mAh$. The length of a cycle is $0.0016+X$ hours. The whole duration in hours will be $1500/(0.12+0.01X) \cdot (0.0016+X)$  and with $X \geq 0.73$ hours, we meet requirement \textbf{R4}, namely a duration of 1 year. 

Note that the smart contract gets in input an unsigned int made of 32 bytes, so we can  pack up to 32 readings (they need 1 byte), into a single call, thus reducing the overall number of required transmissions. 

\section{Conclusions and Future Work}

\todo[inline]{Discutere focus su ex-situ e limii di ex-situ. Anche se solo conservazione la distribuzione implica coinvolgimento}

The workflow  for the conservation of orthodox seeds in seed genebanks \cite{faostandpract}  involves several phases: acquisition, drying and storage, seed viability monitoring and testing, characterization and evaluation, documentation, distributions and safety duplication.

In this paper we limit our attention to the storage and starting from the observation that even SGSV, the most renowned back-up seed facility, has limited redundancy in the availability of some seeds samples we lay out a vision for Distributed Seed Storage (DSS), a novel shared economy concept for long-term storage of seeds in domestic freezers. This concept aims to contribute to the fundamental goal of preserving seed diversity through the employment of IoT to monitor the correct preservation of seeds and Blockchain technologies to distribute incentives to encourage user participation. A modern approach, should foresee the active participation of citizens beyond the simple storage. Nevertheless, we believe the sharing economy principles behind our concept might represent a first step in this direction.

The PoC proves the feasibility of the main technical components of our vision and a future collaboration with fridge producers, might support the development of new products supporting the DSS functionality since the design, also on industrial refrigerators. The PoC also allowed us to identify the priorities to be investigated in future work as summarized below. 

The monitoring activity confirms that acceptable subzero conditions are achievable in domestic freezers, however to what extent we can better approach the ideal conditions prescribed in Standard 4.2.3~\cite{faostand} and the implications on seed conservation of the observed fluctuations, need further investigation. 
While the current board proves the feasibility of the proposed approach, the power consumption can still be optimized and more energy-efficient custom boards, such as the trigboard\footnote{\url{https://trigboard-docs.readthedocs.io/en/latest/}}, can significantly improve the performance. The PoC relies on periodic transmission of the observed conditions, the use of an alternative approach based on the communication of abnormal events only, possibly based on Ultra Low Power Coprocessor Programming, should be evaluated. In this work, a simple lottery incentives users' participation. We have to better understand the complexity beyond a distributed network of a significant number of devices, and to design an indexing service to understand and manage the distribution of the seeds in the network of participants. 

Beyond technical feasibility, we have to evaluate the costs of this solution and to study the most suitable mechanisms to incentivize user participation, not only through more advanced tokenomics mechanisms, but also by leveraging the monitoring system for seeds to implement added-valued services for users, such as the notification of dangerous conditions for the correct conservation of the food in the fridge.     

Facilitating easy and routine access to crop genetic resources is a fundamental goal of the Multilateral System \cite{MLS}. These resources may include local seed collections preserved in small refrigeration units within research laboratories, national seed banks maintained by government institutions, or comprehensive collections held by international research centers that encompass all known varieties of a given crop. Our goal is to explore how the paper's proposed participative approach can be integrated into the MLS. This includes determining how to regulate the exchanges involving the new category of users sharing their fridges, utilizing the Standard Material Transfer Agreements (SMTAs).

\section{Acknowledgments}
This version takes into account the review by Nelizsa Jamora of the original paper (see \url{https://www.qeios.com/read/NAM5SG/pdf})


\bibliographystyle{ACM-Reference-Format}
\bibliography{seed}
\end{document}